# EXTERNAL-CAVITY DESIGNS FOR PHASE-COUPLED LASER DIODE ARRAYS

G. Lucas-Leclin[a], D. Pabœuf[a], P. Georges[a], N. Michel[b], M. Calligaro[b], M. Krakowski[b],
J. J. Lim[c], S. Sujecki[c], E. Larkins[c]

a Laboratoire Charles Fabry de L'Institut d'Optique, CNRS, Univ Paris-Sud, France
b Alcatel-Thales III-V Lab, France
c The School of Electronic and Electrical Engineering, University of Nottingham, United Kingdom

## Introduction

High-brightness single laser diodes based on the taper design have demonstrated output powers of a few Watts with a single transverse mode operation. The use of arrays of such lasers results in a further increase of the laser power, but with the drawback of a loss in the spatial brightness which is reduced as the filling factor of the array. To overcome this limitation, coherent operation of the laser array does improve the brightness. The purpose of this work is to propose and evaluate passive coherent combining architectures, based on external cavity designs. Coherence of the entire laser bars is induced by the diffractive coupling between the emitters within the external cavity. Two different configurations will be described, the first one is based on the Talbot self-imaging effect [1] and the second one consists in the angular filtering of the laser emission [2].

## Description of the experimental configurations

We use arrays of index-guided, tapered laser diodes emitting around 975 nm. The active layer consists of a strained GaInAs quantum well embedded in a large optical cavity. The lateral structure of the emitters is a tapered ridge with a narrow angle (< 1°) and an overall cavity length of 2.5 mm [3]. The slow axis emission width of each emitter is $2w = 30$ µm on the front facet. The tapered design allows single transverse mode operation together with high power emission thanks to the widened active section. The rear facet is high-reflection coated (R > 90 %). The antireflection coating ($R < 10^{-3}$) on the front facet of the laser array prevents self-running operation without external cavity. The array is collimated in the fast axis only with a high NA cylindrical lens (FAC). Finally volume Bragg gratings (VBG) have been utilized as the output coupler of the external cavities. We have taken benefit of their strong spectral filtering properties to stabilize the emission wavelength, as well as their angular selectivity.

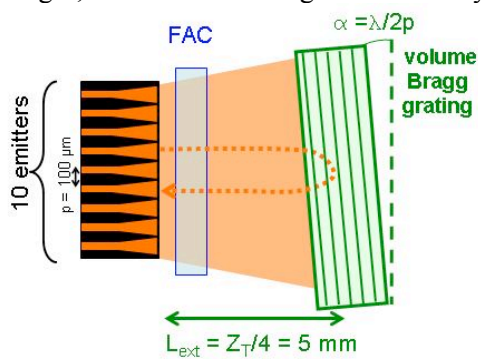
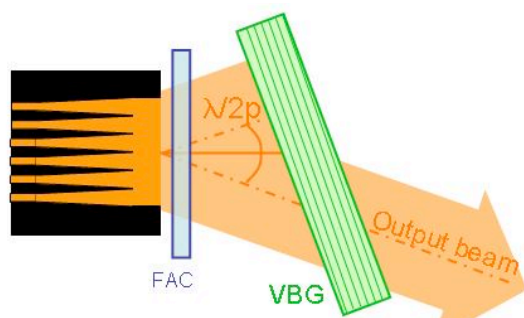

Figure 1 : Talbot extended-cavity diode laser array with a volume Bragg grating

Figure 2 : Angular-filtering extended-cavity design for high fill-factor array

## Numerical modeling of the external cavity operation

A simple theoretical model has been used to predict the modal properties of the extended-cavity lasers. We consider a bar of N single-mode emitters, equally spaced with a pitch p; the array transverse modes are the N eigenmodes of the classical laser equation at steady-state in the external cavity (1), where E is a vector which describes the optical field from each emitter at the front facet, and κ is a N × N matrix, whom (m,n) coefficient is the coupling factor from emitter n towards emitter m. The feedback from the external cavity takes into account the free-space propagation within the external cavity and any angular or spatial filtering that may be added into the cavity (by a volume Bragg grating for example).

$$r_0 r \, e^{2i\varphi} e^{2gL} \{\kappa_{mn}\} \times \vec{E} = \vec{E} \qquad (1)$$

The mode E with the highest eigenvalue will be favoured by the external cavity. This model allows predicting the near-field and far-field profiles of the transverse array modes.

In addition, a Fox&Li iterative approach has been implemented to take into account the propagation and amplification of light inside the tapered semiconductor emitters. It enables us to have an accurate analysis of the propagation losses inside the cavity as well as a fine description of the array transverse modes. It has also been used to calculate the L(I) characteristic [4].

## Results

In the Talbot external-cavity setup (Fig.1), a bar of $N = 10$ lasers with a pitch $p = 100$ µm has been used; the external cavity length is then $p^2/2\lambda = 5$mm. A reflective VBG is used as the external mirror. It is designed to reflect 40% at 976 nm with a spectral bandwidth $\Delta\lambda = 0.3$ nm. A threshold of 0.9 A and a power of 1.7 W at an operating current of 3.9 A have been obtained for the in-phase mode operation of the laser array **Erreur ! Source du renvoi introuvable.**. Thanks to the VBG, each emitter is strongly locked to the Bragg wavelength whatever the operating current and the temperature. It results in a narrow spectrum for the whole bar, below 0.1 nm at 3 dB. We have also demonstrated experimentally that the VBG actually favours the phase-locked operation of the laser bar.

In the angular filtering extended-cavity configuration (Fig.2), a bar of $N = 6$ adjacent tapered lasers with a pitch $p = 30$ µm has been used. No coupling between emitters is observed without the extended-cavity. A reflection VBG with a reflectivity $R \geq 99\%$ at $\lambda_B = 979$ nm reflects the output beam at the angle of $+\lambda/2p$ to select the out-of-phase mode. The angular selectivity of the grating is about $\Delta\theta_R = 35$ mrad. A narrow peak centred at $-\lambda/2p$ and containing up to 50% of the total output power, with a $M^2$ parameter $< 2$ is then observed in the far-field. Furthermore the laser spectrum is locked to the Bragg wavelength $\lambda_B$ resulting in a <0.1 nm-wide stabilized line. The total output power reaches 700 mW at 3 A.

## Conclusion

Both configurations have been extensively studied theoretically as well as experimentally. We have obtained a stable coherent operation of the laser arrays, on a single transverse array mode. The main advantage of the angular filtering cavity is here that it is perfectly adapted to the high filling-factor laser bar, and then provides directly a nearly diffraction-limited output beam; nevertheless these bars have limited output powers because of the strong heating of the array. On the other hand, Talbot external cavities may be used with very different kind of bars, in simple extended-cavity designs. Finally, the use of the filtering properties of VBG has been demonstrated to efficiently control the laser emission.

## Acknowledgements

The authors thank the European Community for financial support under the www.BRIGHTER.eu program (IP 035266). D. Paboeuf acknowledges the funding of his PhD by the French Délégation Générale de l'Armement.